# Recrossing and tunnelling in the kinetics study of the
# OH + CH$_4$ → H$_2$O + CH$_3$ reaction.

Yury V. Suleimanov[a,b,*] and J. Espinosa-Garcia[c,*]

[a] *Department of Chemical Engineering, Massachusetts Institute of Technology, 77 Massachusetts Ave., Cambridge, Massachusetts 02139, United States*

[b] *Computation-based Science and Technology Research Center, Cyprus Institute, 20 Kavafi Str., Nicosia 2121, Cyprus*

[c] *Departamento de Química Física, Universidad de Extremadura, 06071 Badajoz, Spain*

* Corresponding authors: y.suleymanov@cyi.ac.cy, joaquin@unex.es

**Abstract**

Thermal rate constants and several kinetic isotope effects were evaluated for the OH + CH$_4$ hydrogen abstraction reaction using two kinetics approaches, ring polymer molecular dynamics (RPMD), and variational transition state theory with multidimensional tunnelling (VTST/MT), based on a refined full-dimensional analytical potential energy surface, PES-2014, in the temperature range 200-2000 K. For the OH + CH$_4$ reaction, at low temperatures, T = 200 K, where the quantum tunnelling effect is more important, RPMD overestimates the experimental rate constants due to problems associated with PES-2014 in the deep tunnelling regime and to the known overestimation of this method in asymmetric reactions, while VTST/MT presents a better agreement, differences about 10%, due to compensation of several factors, inaccuracy of PES-2014 and ignoring anharmonicity. In the opposite extreme, T = 1000 K, recrossing effects play the main role, and the difference between both methods is now smaller, by a factor of 1.5. Given that RPMD results are exact in the high-temperature limit, the discrepancy is due to the approaches used in the VTST/MT method, such as ignoring the anharmonicity of the lowest vibrational frequencies along the reaction path which leads to an incorrect location of the dividing surface between reactants and products. The analysis of several kinetic isotope effects, OH + CD$_4$, OD + CH$_4$, and OH + $^{12}$CH$_4$/$^{13}$CH$_4$, sheds light on these problems and confirms the previous conclusions. In general, the agreement with the available experimental data is reasonable, although discrepancies persist, and they have been analysed as a function of the many factors affecting the theoretical calculations: limitations of the kinetics methods and of the potential energy surface, and uncertainties in the experimental measurements. Finally, in the absence of full-dimensional quantum mechanics calculations, this study represents an additional step in understanding this seven-atom hydrogen abstraction reaction.

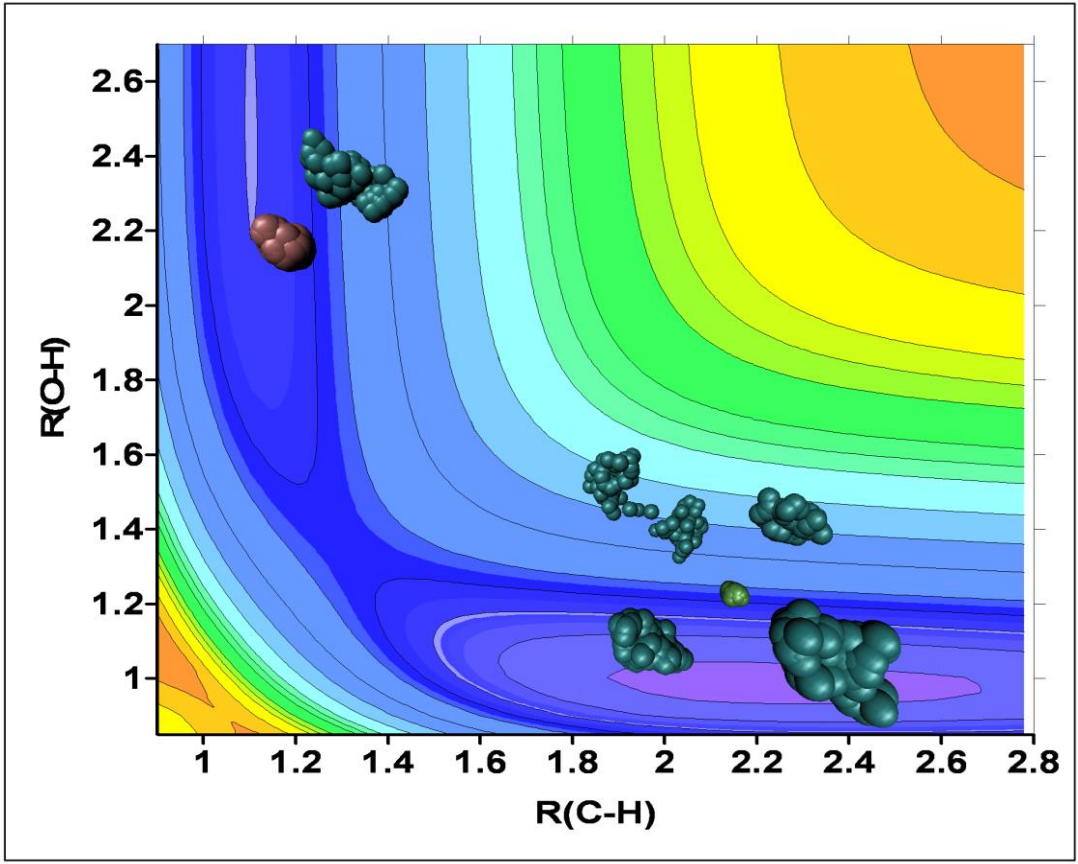

TOC graphic

# 1. Introduction

The combination of potential energy surfaces (PES), which describes the nuclear motion and kinetics models represents a useful strategy for calculating rate constants and related properties: tunneling, recrossing and kinetic isotope effects (KIE).

PESs have been generated by using different techniques, from the description of the reactive system using electronic structure calculations (energy, gradients and Hessians) carried out "on the fly", sometimes named "direct dynamics"; to the development of mathematical expressions, i.e., algorithms which provide the energy of the system (and sometimes the gradient) for any geometry of the reactive system. As for the kinetics models, different theories have also been developed, each with their advantages and limitations. These include variational transition state theory with multidimensional tunnelling (VTST/MT),[1-6] quasi-classical trajectory (QCT),[7] quantum mechanical (QM) (especially useful in small reactive systems),[8] multiconfiguration time-dependent Hartree (MCTDH),[9,10] and, more recently, ring polymer molecular dynamics (RPMD).[11-13]

The OH + CH$_4$ hydrogen abstraction reaction is of great importance at low and high temperatures: at low temperatures it constitutes the major process for the removal of methane in atmospheric chemistry, in competition with the O($^1$D) and Cl($^2$P) reactions; and at high temperatures because of its importance in combustion. The thermal rate constants for the title reaction have been reported experimentally using a wide variety of techniques,[14-36] and the two most recent evaluations[33,36] recommend the following expressions, in cm$^3$molecule$^{-1}$s$^{-1}$:

$$k(T) = 1.36.10^{-13} \ (T/298)^{3.04} \ \exp(-7.65/RT) \qquad T=195\text{-}1234 \ K \qquad (1)$$

and

$$k(T) = 4.16.10^{-13} \ (T/298)^{2.18} \ \exp(-10.24/RT) \qquad T=195\text{-}2025 \ K \qquad (2)$$

Despite being one of the most studied gas-phase reactions, these experimental evaluations still present some discrepancies. At low temperatures (195 K), while Bonard et al.[33] recommend 3.35.10$^{-16}$, Srinivasan et al.[36] report 2.99.10$^{-16}$ cm$^3$molecule$^{-1}$s$^{-1}$; and at high temperatures (1000 K), 2.15.10$^{-12}$ versus 1.70.10$^{-12}$ cm$^3$molecule$^{-1}$s$^{-1}$. Such a small difference reflects the experimental difficulties in the rate constant measurements. Given the interest of this reaction, different experiments[14,28,32,35, 37-40] have also reported kinetic isotope effects (KIEs).

Due to its atmospheric and combustion interest, the kinetics of the title reaction has also been widely studied theoretically using different methods.[41-60] Most of these studies used potential energy surfaces constructed "on the fly" based on *ab initio* calculations. We use an alternative strategy, developing analytical PESs, which provides analytically the energy and the first energy derivatives of the reactive system, which represents a computational advantage. In 2000, our lab developed[53] the first analytical PES for the title reaction, PES-2000, which is symmetric with respect to the permutation of the four methane hydrogen. In recent years, PES-2000 has been used as a testing bench for different kinetics approaches.[53,59,60] In general, the agreement with the available experimental data is good. However, the major limitation of PES-2000 is its semiempirical character, i.e., in the fitting procedure theoretical and experimental information was used. To correct this serious deficiency, which limits the applicability of the surface, we have recently reported a new surface, PES-2014,[61] which is fitted exclusively to high-level *ab initio* calculations. This surface is invariant with the methane hydrogen permutation, and not only provides the energy but also the gradients, which represents an advantage over other surfaces.

In the present paper, we report a kinetic study of the title reaction based on PES-2014 by using two different approaches: ring polymer molecular dynamics and variational transition state theory with multidimensional tunnelling. Our main aims are, first to analyse different effects affecting the thermal rate constants, tunnelling and recrossing; second, to analyse their influence in comparison with experiments and, finally, to test the two kinetics approaches, since quantum-mechanical methods for full-dimensional polyatomic systems reactions (in this case seven atoms) are practically forbidden today. This paper is structured as follows: Section 2 briefly summarizes the potential energy surface and the computational aspects of the RPMD and VTST/MT theories; the results of the present calculations are discussed in Section 3, and finally, the conclusions are presented in Section 4.

## 2. Potential energy surface and computational details.

**a) Potential energy surface**. In 2014 we developed a new full-dimensional analytical potential energy surface, PES-2014, describing the $OH + CH_4 \rightarrow H_2O + CH_3$ gas-phase reaction,[61] which improves the previous one, PES-2000,[53] which was also developed in our lab. The functional form and the fitting procedure were widely developed in the original work[61] and they are briefly exposed here to avoid unnecessary repetitions.

Basically, PES-2014 is a valence bond-molecular mechanics (VB-MM) surface, with 35 adjustable parameters, which were fitted to high-level *ab initio* calculations,

$$V = V_{stret} + V_{val} + V_{op} + V_{H2O} \qquad (3)$$

where these functions represent, respectively, the Londong-Eyring-Polanyi stretching terms, depending on 12 parameters; the harmonic bending terms, depending on 16 parameters; the anharmonic out-of-plane potential, which depends on 4 parameters, and the potential describing the water product, which depends on 3 parameters. In addition, this surface includes a series of switching functions allowing the relaxing from tetrahedral methane structure to planar methyl radical and from hydroxyl radical to water product. PES-2014 provides analytically the energy and the gradients. This surface has a barrier height of 6.4 kcal mol$^{-1}$, reproducing recent and accurate high-level *ab initio* calculations,[62,63] 6.1-6.7 kcal mol$^{-1}$, and an exothermicity of -13.3 kcal mol$^{-1}$, reproducing *ab initio* calculations,[62] -13.5 kcal mol$^{-1}$. In addition, it presents stabilized intermediate complexes in the entrance and exit channels, and in general it reproduces the reaction path topology from reactants to products.

However, in the fitting procedure we focussed the attention on the reaction path valley, the region represented by the minimum energy path and orthogonal generalized normal-mode coordinates, while the region with higher energies outside this valley is not expressly considered. Sometimes, this region is known as reaction swath, which includes geometries and energies on the concave side of the minimum energy path. Nevertheless, in the present paper we have found that this region can be accessed by tunnelling paths when reaction path curvature is large, which is the case for this heavy-light-heavy reaction. Although one expects that the VB-MM algorithm is able to extrapolate the potential energy for regions not included in the fitting, i.e., higher energies, as it was observed in a recent work[64] for the O($^3$P) + CH$_4$ reaction, where experimental measures are reproduced at 64 kcal mol$^{-1}$; obviously this is not a general tendency and in general the accuracy will be limited. Figure 1 plots the PES-2014 contour plot (upper panel), where the large curvature tunnelling region (reaction swath) is highlighted; and the reaction path (lower panel), i.e., the minimum energy path starting from the saddle point structure. We return to this point in the discussion of results (Section 3a).

**b) VTST/MT approach**. The thermal rate constants were calculated using the canonical variational theory,[65,66] CVT, which locates the dividing surface along the reaction coordinate(s) at the maximum of the free energy of activation, $\Delta G^{GT,o}$ $(T, s^{*,CVT})/K_B T$),

$$k^{CVT}(T) = \sigma \frac{K_B T}{h} K^o \exp\left[-\Delta G(T, s^{*,CVT})/K_B T\right] \qquad (4)$$

where $\sigma$ is the symmetry factor (12 for this reaction), $K_B$ is the Boltzmann constant and $K^o$ is the reciprocal of the standard-state concentration, 1 molecule $cm^3$. The rotational partition functions were calculated classically, and the $^2\Pi_{1/2}$ excited state of OH (140 $cm^{-1}$)[67] is taken into account in the reactant electronic partition function. All vibrational modes are treated as separable harmonic oscillators using curvilinear coordinates,[68-70] with exception to the lowest vibrational mode along the reaction coordinate which corresponds to practically free rotation of the OH bond on the new $(H_3C)H'$-O bond formed. It is treated using the hindered rotor model (RW model)[71] to include anharmonicity.

In this heavy-light-heavy mass combination reaction, the tunnelling contribution plays a very important role at low temperatures and this contribution was calculated using the large curvature tunnelling (LCT) method, with three semiclassical approaches: LCG3, LCG4 and LAT.[66,72,73,74.] The LCT methods need information on the reaction swath, and are computationally expensive. The LCG4 method improves the previous LCG3 approach by including a nonquadratic correction, i.e., anharmonicity, in the nonadiabatic region; while the LAT method, recently developed for polyatomic reactions,[74] is the most sound and complete of the series.

Finally, we note that although VTST/MT is an approximate theory, it is a well-validated approach to treat zero-point energy, recrossing and tunneling.[75] This theory has been tested against accurate quantum rate constants for 74 atom-diatom reactions,[76] finding that when using the LAT approach, the rate constants are generally in good agreement, with differences of 30-35% and 20-25% when anharmonicity is not and is included, respectively. This result shows the influence of the anharmonicity, which seems small. More recently,[74] for the polyatomic H + $CH_4$ reaction, a difference about 10% at 200 K was reported. All kinetics calculations were performed with the POLYRATE-2010 code.[77]

**c) RPMD approach**.

Ring polymer molecular dynamics is a recently developed full-dimensional dynamics approximation which is based on the classical isomorphism between the quantum system and the series of its classical copies placed in the necklace forming ring structure in an extended phase space with a harmonic interaction between the neighbouring beads.[78] The classical evolution of this ring polymer is used to approximate the real-time dynamics in the original quantum system.[11-13] It was recently found that such an approximation provides very accurate and reliable estimates for thermal rate constants in systems with various dimensionalities and energy profiles and in different temperature regimes:[60,79-93] RPMD rate is exact in the high-temperature limit (i), reliable at intermediate temperatures (ii), more accurate than other methods in the deep quantum tunneling regime, within a factor of 2-3 of the exact quantum results (iii), captures perfectly the zero-point energy effect (iv), also provides accurate estimates for reactions of barrierless type (v). RPMD provides systematic and consistent performance − it overestimates thermal rates for asymmetric reactions and underestimates them for symmetric reactions in the deep tunneling regime.[86,94] (Note that the zero-point energy effect along the reaction coordinate must be taken into account when assigning the reaction symmetry.)

All RPMD calculations reported here used the RPMDrate developed by Suleimanov and co-workers and the remaining details of the computational procedure can be found in the RPMDrate manual.[83] We use the same parameters as those in our previous study[60] of the title reaction with the older PES-2000.

**3. Results and discussion**

**a) OH + CH$_4$ → H$_2$O + CH$_3$ reaction.** We begin by analysing the rate constants for the OH + CH$_4$ reaction because more accurate and recent experimental measures are available for comparison. Figure 2 plots the thermal rate constants using the VTST/MT and RPMD kinetics methods on PES-2014, with experimental data, in the temperature range 200-2000 K (VTST/MT) and 200-1000 K (RPMD).

Taking the experimental values as a reference, we analyzed both the kinetic method and the surface. The VTST/MT method reproduces the experimental behaviour in the complete temperature range, with differences between 3% and 27% in the common temperature range (200-2000 K), ~20% at low temperatures (T ≤ 300 K) and

~10% at high temperatures (T ≥ 1000 K). The differences for the RPMD method are larger, ~45% at T ≤ 300 K, overestimating the experiment, and 85% at T = 1000 K, underestimating it.

We begin by analysing the behaviour at low temperatures, where the quantum mechanical tunnelling effect plays the most important role. For this reaction the crossover temperature is $T_c = 370$ K. It has been shown that RPMD overestimates the rate constants for asymmetric reactions[86,94] and this behaviour is also observed in the title reaction. Previous experience on hydrogen abstraction reactions[60,79-81,83,85,86,88,89,91] shows that RPMD overestimates the exact rate constants by a factor of about 2-3 in the tunnelling regime. In the present reaction, however, RPMD rate constant is about four times larger than the experimental values at 200 K. Given that RPMD is largely checked in very different reactive systems, this implies problems related with the accuracy of the PES-2014 for this seven atom system. Thus, although the barrier height, 6.4 kcal mol⁻¹, and the topology of the reaction path near the transition state region obtained with PES-2014 match high-level *ab initio* information,[61] (see Figure 1), the accurate description of zones far this region, which dominates the tunnelling regime, were not checked.[61]

Now we perform a deeper analysis of this issue. Figure 3 shows the PES-2014 and *ab initio* contour plots in the large curvature tunnelling region, where the red straight lines schematically simulate tunnelling paths in this region, and they are included simply as a visual guide for readers making the comparison between both surfaces easier to visualise. It is observed that these paths "visit" regions of higher energies in the case of the *ab initio* surface, i.e., PES-2014 is less repulsive and thinner than the *ab initio* surface. As a consequence, PES-2014 will provide larger tunnelling effects, and this is the behaviour found when the RPMD theory is used. Therefore, a part of the overestimation of the tunnelling by RPMD is due to limitations of the PES-2014 description of the large curvature tunnelling region.

On the other hand, in this temperature regime it is important to note that the tunnelling effect in the VTST/MT method is calculated as a multiplicative factor using semiclassical expressions, which depend on the approach used. Thus, at 200 K, while the LCG3 method gives a factor 14.6, the LCG4 and LAT methods give, respectively, 11.2 and 11.3, i.e., for the title reaction, dependence on the tunnelling approach is small. However, due to the multidimensional nature of the quantum tunnelling effect, the LCT semiclassical approaches included in the VTST/MT theory can be often unreliable in this tunnelling regime.[60,91] Thus, given the PES-2014 limitations in this region and the semiclassical character of the tunneling corrections, the VTST/MT-experiment agreement found for this reaction can be due to error cancellation.

Next, the high-temperature regime is analysed. Given that RPMD is exact at high temperatures the underestimation at 1000 K seems to show that the discrepancy is associated with the VTST/MT method and/or PES-2014. In the latter case, this implies that the barrier height should be lower. However, given this barrier height reproduces high-level values, the discrepancy seems to be associated with the VTST/MT approach and more specifically with the evaluation of recrossing effects. At this point, it is important to note that now the differences RPMD-VTST/MT are smaller, by a factor of 1.5, i.e., about three times lower than in the low temperature regime. In the RPMD theory the recrossing effect is obtained by analysing the long-time limit of the corresponding ring-polymer correlation functions.[79,80] At 1000 K the RPMD transmission coefficient is very small, 0.44, indicating high recrossing, which is an expected behaviour in the heavy-light-heavy reactions.[60,79,85,87,89,91] In the VTST/MT theory the recrossing effect is measured as the ratio between CVT(s = s*) and TST(s = 0, saddle point) rate constants. It is also known as "variational effect" and measures the effect of the shift of the maxima of the free energy curve, s*, from the saddle point, s = 0. At 1000 K, the ratio CVT/TST is 0.48 (the coincidence between the two factors is merely casual). Compared with the exact RPMD rate constant, this indicates that the recrossing effect obtained with the VTST method is underestimated. This deficiency of the VTST approach is related to the location of the dividing surface between reactants and products, which is intimately associated with the decrease of the low vibrational frequencies as one leaves the saddle point.[58] Note that RPMD is immune to this problem because it is independent of the location of the dividing surface.[13] In the VTST theory the location of the dividing surface depends on the choice of the coordinate system used and on the harmonic/anharmonic treatment of the vibrational frequencies. With respect to the first factor, while the maximum of the free energy curve in the Cartesian coordinates is located at s* = -0.5478 bohr, in the curvilinear coordinates it is located at s* = -0.3030 bohr, i.e. closer to the saddle point. This leads to small differences in the recrossing factors, 0.45 and 0.48, respectively. In addition, at this point it is important to note that while the curvilinear coordinates yield real vibrational frequencies along the reaction path, the Cartesian coordinates yield some imaginary frequencies, which is a highly undesirable fact. With respect to the second factor, we found previously[88,95] that when the lowest vibrational modes are considered anharmonic they fall sharper along the reaction path from the saddle point than when they are treated as harmonic oscillators. This behaviour leads to smaller variational effects and less recrossing.

To estimate the influence of these effects on recrossing, PES and anharmonicity, we have performed quasi-classical trajectory (QCT) calculations at high temperatures, T = 1000 K, where tunnelling is negligible. The QCT initial and final conditions were already detailed in our previous paper,[61] and they are not repeated here. The main difference is that while in that paper we use a fixed collision energy, here we use a fixed temperature. Then, we obtain rate constants of 13.9, 10.2 and $9.26.10^{-13}$ cm$^3$ molecule$^{-1}$ s$^{-1}$ using VTST/MT, QCT and RPMD, respectively. So, taking the RPMD value as accurate, the VTST/MT and QCT methods overestimate it by 50 and 10 %, respectively. Since the three methods use the same PES-2014, this shows that the PES is responsible for the 10% difference, and the location of the dividing surface in the VTST/MT theory is responsible for the other 40 %. Obviously, this is a semi-quantitative approach, because the PES and the dividing surface minimizing the recrossing effect are interrelated issues. In addition, given the zero-point energy violation problem in QCT calculations, its result represents a upper limit of the rate constant, and so it approaches to the RPMD result, which is free of this problem.

**b) Kinetic isotope effects.** To understand the influence of tunnelling and recrossing and to shed more light on the kinetics of the title reaction, the KIEs represent a powerful tool. The OH + CD$_4$ → HOD + CD$_3$ isotopic variant transfers a deuterium atom instead of the H atom, and therefore it presents less tunnelling effect. The crossover temperature for this reaction is T$_c$ = 269 K. The rate constants in the common temperature range 200 -1000 K are plotted in Figure 4 for the VTST/MT and RPMD methods. Now both methods agree at low temperatures, with a factor of 1.13 at 200 K, against a factor of 4.5 for the OH + CH$_4$ reaction, while at high temperatures the difference is similar, 2.0 versus 1.5. These results confirm the previous conclusions. Thus, at low temperatures, T < T$_c$, where tunnelling effect dominates, and for the deuterated reaction this effect plays a minor role, both methods lead to similar results, which seems to confirm that the discrepancies in the OH + CH$_4$ reaction (Fig. 2) are associated to deficiencies in PES-2014 describing the deep tunnelling regime. At high temperatures the recrossing effect dominates, which is better captured by the RPMD method. The anharmonic effects in the VTST/MT theory are more important for the deuterated methane and therefore the difference is a bit larger.

The $CH_4/CD_4$ KIEs are plotted in Figure 5 in this temperature range and compared with the previous experimental measures.[28,32] VTST/MT method underestimates experimental KIEs by a factor of ~ 2 in the common temperature range, while RPMD method presents a good agreement with experiments at high temperatures, but strongly overestimates them in the deep tunnelling regime.

At this point two issues must be highlighted. First, it is important to note that while the experimental rate constants for the $OH + CH_4$ reaction have been recently revised, 2005 (Ref. 36), the experimental data for the $OH + CD_4$ reaction are older, 1997 (Ref. 32), and they have not been recently revised. Given the good agreement obtained for the $OH + CH_4$ reaction using the VTST/MT method in all temperature ranges, except at low temperatures (Fig. 2) it could be due to error cancellation, this seems to indicate that the experimental data for the $OH + CD_4$ reaction must be experimentally re-evaluated. Second, since KIEs are less dependent on the accuracy of the PES, they represent a direct test on the performance of each method. Thus, these results show that while RPMD captures the recrossing effects at high temperatures better than VTST/MT, the RPMD overestimation at low temperatures is due to deficiencies of the PES-2014 surface and the known systematic overestimation of this method in asymmetric reactions.

A second test of these effects was performed by studying the isotopic variant $OD + CH_4$ reaction, where again a hydrogen atom is transferred and the tunnelling effect plays an important role at low temperatures (crossover temperature is $T_c = 370$ K). Figure 6 plots VTST/MT and RPMD rate constants in the temperature range 200-1000 K. The variation of the rate constants with temperature presents similar behaviour to the $OH + CH_4$ reaction. Thus, at low temperatures, $T < T_c$, the RPMD method seems to overestimate the VTST/MT values, with differences of a factor of ~ 5 at 200 K, while the opposite tendency is found at high temperatures, with differences of ~ 0.7 at 1000 K. These results confirm previous conclusions for both low and high temperature regimes.

The OH/OD KIEs in the range 200 - 1000 K are plotted in Figure 7 along with experimental measurements for comparison. In the common temperature range (250-416 K) all KIEs present an "inverse" behaviour, i.e., smaller than 1, but while the VTST/MT method shows good agreement with the experiment, with differences ≤ 10%, the RPMD method underestimates the experiment, ~ 25-30. While the agreement obtained with the VTST/MT method can be due to cancellation of the tunnelling effect in both isotopic reactions, the RPMD behaviour is not easy to explain, because one should expect that the systematic tunnelling overestimation cancels out in these reactions. As it was previously noted, since the KIEs are less dependent on the PES parameters, these results show the limitations included in both methods, and/or uncertainties in the experimental data.

Finally, the $^{12}$C/$^{13}$C KIEs are analysed, which represent a severe test of the theoretical tools used, because many factors are involved. These KIEs have been previously studied by several groups.[35,58,96] They pointed out that these KIEs are strongly dependent on the methods used to treat the torsional anharmonicity of the lowest-frequency vibrational mode, and less dependent on the PES used. Table 1 lists the KIEs obtained in this work along with the experimental values. While the VTST/MT method obtains "normal" KIEs, i.e., larger than 1, reproducing the experimental evidence, the RPMD method obtains "inverse" KIEs, i.e., < 1. However, VTST/MT overestimates the experiment by ~ 5% while RPMD underestimate it by a very small amount of less than 1% which is even below convergence in the RPMD calculations. Such excellent agreement between experiments and RPMD is due to the fact that the latter is a full-dimensional approach which treats all degrees of freedom on an equal.

To analyse the overestimation of 5% in the VTST/MT method we perform a factorization analysis in Table 2, comparing the results with recent high-level *ab initio* calculations.[58] For direct comparison, T = 296 K and small curvature tunnelling (SCT) approach are used. Note that the large curvature approach (LCT) is only possible when an analytical PES is used, and this is only available from the PES-2014 surface. The KIEs factorize as[97]

$$\text{KIE}(^{12}\text{C}/^{13}\text{C}) = \eta_{\text{trans}} \cdot \eta_{\text{rot}} \cdot \eta_{\text{vib}} \cdot \eta_{\text{pot}} \cdot \eta_{\text{tunn}} \tag{5}$$

where $\eta_{\text{trans}}$, $\eta_{\text{rot}}$ and $\eta_{\text{vib}}$ are the ratio of translational, rotational and vibrational partition functions, respectively, $\eta_{\text{pot}}$ is from the potential energy curve at the generalized transition state, and $\eta_{\text{tunn}}$ is the tunnelling contribution, where the SCT approach is used for a direct comparison. The *ab initio* PES and the analytical PES-2014 show excellent agreement for the factors $\eta_{\text{trans}}$, $\eta_{\text{rot}}$, $\eta_{\text{pot}}$ and $\eta_{\text{tunn}}$, with the largest difference in the $\eta_{\text{vib}}$ contribution. This difference, 1.003 versus 0.973, which is small, is responsible for the difference in the final KIEs.

Let's analyse in more detail the $\eta_{vib}$ contribution for the PES-2014. Table 3 lists the vibrational frequencies at the bottleneck and the reactants for the $^{12}CH_4$ and $^{13}CH_4$ systems. The largest difference in the reactants is 11 cm$^{-1}$, and in the generalized transition state it is 13 cm$^{-1}$. This very small difference (close to the spectroscopic accuracy) leads to strong dependence of the $^{12}C/^{13}C$ KIEs on the coordinate system used to calculate the frequencies and on the method used for the torsional anharmonicity of the lowest frequency. We have analysed the first factor, and in addition we performed some tests with other coordinate systems: Cartesian and non-redundant. In both cases, imaginary frequencies appear in the reaction path, and so they were discarded. Now we analyse the second factor. The partition function of this lowest vibrational mode in the generalized transition state is 2.323 and 2.333 for the $^{12}C$ and $^{13}C$ isotopes, respectively. This represents a factor of 0.996. Another factor is the tunnelling contribution. While the direct dynamics method (where energy, gradient and Hessian are calculated when needed by electronic structure theory) only permits the small curvature tunnelling (SCT), the analytical PES-2014 permits the calculation of large curvature tunnelling (LCT) and therefore, the microcanonical optimized multidimensional tunnelling ($\mu$OMT) approach (at each energy the best estimate is the larger of SCT and LCT probabilities). Using PES-2014 the SCT $^{12}C/^{13}C$ factor is 1.010 (Table 2), while the $\mu$OMT $^{12}C/^{13}C$ factor is 1.008, i.e., the difference is practically negligible. Therefore, these factors explain the disagreement theory/experiment for the $^{12}C/^{13}C$ KIEs, and the other source of error, the potential energy surface, seems to play a minor role.

**c) Comparison with previous kinetics works.** In the present work, only two kinetics methods, VTST/MT and RPMD, have been tested using PES-2014, and there is no theoretical information about other kinetics approaches. However, in 2012 and 2013 two kinetics studies appeared[59,60] in which the VTST/MT approach was compared with the quantum instanton (QI) and the RPMD approaches, although our older PES-2000 surface was employed.[53] At high temperatures (T = 1000 K), the three methods reproduced the experimental rate constants for the OH + CH$_4$ reaction, with differences $\sim \leq 20\%$, showing that the recrossing effects are reasonably well captured in all cases. However, at low temperatures, in the deep tunnelling regime (T $\leq$ 250 K), the differences are larger. At 250 K, where comparison is possible, while the VTST/MT method presents the better agreement with experiment, which is merely fortuitous because the rate constants were used in the fitting procedure; the RPMD and QI methods overestimates it by factors of ~ 4 and ~ 1.5, respectively. The last two methods are expected to capture better the tunnelling contribution, the most important effect in this temperature regime.[59,60,91] This result suggests that the PES-2000 surface is too thin with respect to accurate *ab initio* calculations, similar to PES-2014. In addition, the computational procedures developed for both RPMD and QI do not require that one calculate explicitly the absolute quantum mechanical partition function of the reactants or the transition state, avoiding problems of VTST/MT calculations related with the anharmonicity of the lowest vibrational modes.

Another interesting aspect is the comparison of the different kinetics methods for the $CH_4/CD_4$ KIEs, which are less sensitive to the accuracy of the PES. For instance, at 250 K, where the comparison with experiment is possible, using PES-2000 the VTST/MT, RPMD and QI values are, respectively, 5.73, 23.47 and 41.11,[53,59,60] as compared to experiment, 10.19.[32] The TST-based methods give contradictory estimates of the KIE, VTST/MT underestimates it by a factor ~ 2, while QI overestimates it by a factors of ~ 4. On contrary, RPMD exhibits expected behaviour. While in the OH + $CH_4$ reaction the tunnelling contribution is very important because a hydrogen atom is transferred, this effect is less important for the OH + $CD_4$ isotopic variant. As a result, the KIE is overpredicted by about factor of ~ 2 at low temperatures due to the anticipated overprediction of the rate for the first reaction by the RPMD theory (by the same factor) due to systematic overestimation of the tunnelling for asymmetric reactions.[60] The QI behaviour is more intriguing. Since the OH + $CH_4$ rate constant is overestimated by only a factor ~ 1.5, this KIE suggests that the OH + $CD_4$ rate constant is underestimated by a factor ~ 2.5. This result does not seem to have much logic, when the same PES-2000 is used, although uncertainties in the experimental results cannot be discarded.

In sum, more accurate theoretical results are necessary (PES and kinetic methods) to solve these discrepancies, especially important in the deep tunnelling regime.

## 4. Conclusions

In this work, we report a kinetics study on the gas-phase OH + $CH_4$ → $H_2O$ + $CH_3$ reaction and some isotopic variants, OH + $CD_4$, OD + $CH_4$ and OH + $^{13}CH_4$, in a wide range of temperatures. Using recently refined full-dimensional analytical potential energy surface PES-2014, we employed two kinetics approaches to calculate thermal rate constants, namely, variational transition state theory with multidimensional tunnelling (VTST/MT) and ring polymer molecular dynamics (RPMD), and compared the results with available experimental measures. This comparison must be performed with caution, because both the PES and the kinetics methods are being simultaneously tested.

At low temperatures, tunnelling effects play the dominant role. In this temperature regime, while RPMD strongly overestimates the tunnelling contribution, VTST/MT reproduces the experiment. In the RPMD case, we have found that this overestimation is related firstly to problems of the surface, it is thinner than the *ab initio* data in reference, producing an artificial overestimation of this quantum effect, and secondly to the known and systematic overestimation of the RPMD method in asymmetric reactions. The good agreement in the VTST/MT case is due to the compensation between the limitations of the surface and the semiclassical nature of the approaches used to estimate the tunnelling effect.

At the opposite extreme, the recrossing effect plays a more important role. In the high temperature regime RPMD theory is exact and the discrepancies with the experiment, underestimation by a factor of 1.8 at 1000 K, are due to limitations in the PES, especially the barrier height, which should be lower. However, given the PES-2014 barrier, 6.4 kcal mol$^{-1}$, matches recent and accurate high-level *ab initio* calculations, 6.1-6.7 kcal mol$^{-1}$, we are within the chemical accuracy ($\pm$ 1 kcal mol$^{-1}$). Lowering artificially the barrier so that RPMD theory matches the experiment, should be equivalent to make semiempirical the PES, which is not the purpose of these authors. On the other hand, the discrepancies with the VTST/MT results arise from problems related with the location of the dividing surface along the reaction coordinate in this theory, especially the anharmonic treatment of the lowest vibrational frequencies. In any case, the differences RPMD-VTST/MT, factor of 1.5 at 1000 K, are smaller than those found at low temperatures, factor of $\sim$ 4.

The analysis of the kinetic isotope effects for some isotopic variants, $CH_4/CD_4$, OH/OD and $^{12}C/^{13}C$, shed more light on these differences, confirming previous conclusions. In the $CH_4/CD_4$ case, at low temperatures, due to limitations of PES-2014, excessively thinner, the RPMD method overestimates the tunnelling contribution, while at high temperatures the VTST/MT approach presents problems related to the location of the dividing surface between reactants and products. In the $^{12}C/^{13}C$ case, we find excellent agreement between RPMD and the experiment while VTST/MT results are slightly higher due to approximations used to treat torsional anharmonicity of the lowest-frequency vibrational mode. Nevertheless, the agreement theory/experiment is generally reasonable for the KIEs studied in this work, and the discrepancies arise from both the theoretical tools (method and potential energy surface) and from experimental uncertainties.

**Table 1**. $^{12}C/^{13}C$ kinetic isotope effects for the OH + CH$_4$ reaction.

| T(K) | VTST/MT | RPMD | Exp.[a] |
|------|---------|------|---------|
| 200  | 1.055   |      |         |
| 250  | 1.043   |      |         |
| 273  | 1.039   |      | 1.005   |
| 293  | 1.037   |      | 1.005   |
| 300  | 1.036   | 0.997|         |
| 353  | 1.031   |      | 1.005   |
| 400  | 1.027   |      |         |
| 500  | 1.023   |      |         |
| 1000 | 1.015   |      |         |

a) Experimental values from Ref. 39.

**Table 2**. Factor analysis of the $^{12}C/^{13}C$ kinetic isotope effects at T=296 K using the VTST/MT method.

| Factor | PES-2014[a] | Ellington et al.[b] |
|---|---|---|
| $\eta_{vib}$ | 1.003 | 0.973 |
| $\eta_{rot}$ | 0.971 | 0.971 |
| $\eta_{trans}$ | 1.047 | 1.047 |
| $\eta_{pot}$ | 0.997 | 1.000 |
| $\eta_{tunn}$ | 1.010 | 1.013 |
| KIE | 1.027 | 1.001 |

a) CVT/SCT calculations using the PES-2014 surface with the RW ($CH_3$-$H_2O$) approach for the torsional motion.

b) Ref. 58. CVT/SCT calculations using the MCG3/3 *ab initio* method with the RW ($CH_3$-$H_2O$) approach for the torsional motion.

**Table 3**. Harmonic vibrational frequencies (cm$^{-1}$) of the generalized transition state(GTS) at s = -0.300 bohr and of the reactants for the $^{12}CH_4$ and $^{13}CH_4$ reactions.

| CH$_4$ reactant | | OH-CH$_4$ (GTS) | |
| --- | --- | --- | --- |
| $^{12}$CH$_4$ | $^{13}$CH$_4$ | $^{12}$CH$_4$ | $^{13}$CH$_4$ |
| 3180 | 3169 | 3803 | 3803 |
| 3180 | 3169 | 3228 | 3215 |
| 3180 | 3169 | 3228 | 3215 |
| 3007 | 3007 | 3077 | 3074 |
| 1546 | 1546 | 1884 | 1877 |
| 1546 | 1546 | 1467 | 1467 |
| 1380 | 1371 | 1422 | 1420 |
| 1380 | 1371 | 1416 | 1411 |
| 1380 | 1371 | 1336 | 1332 |
| | | 1097 | 1090 |
| | | 795 | 794 |
| | | 478 | 477 |
| | | 478 | 476 |
| | | 29 | 29 |

**FIGURE CAPTIONS**

**Figure 1.** PES-2014 contour plot in the proximity of the saddle point (upper panel) and reaction paths of the PES-2014 and *ab initio* surfaces (lower panel), where s<0 and s>0 represent, respectively, the reaction coordinate for the reactant and product regions. R(OH) and R(CH) in Angstrom.

**Figure 2.** Arrhenius plots of the OH + CH$_4$ rate constants computed using PES-2014: CVT/LCG3, blue line; RPMD, black line; experimental from Ref. 36, red line; CUS/LCT-3 using PES-2000 from Ref. 53, green line. Vertical dashed line represents the crossover temperature, T$_c$ = 370 K.

**Figure 3**. Contour plots of PES-2014 (upper panel) and *ab initio* (lower panel) surfaces representing the large curvature tunneling region (see Fig. 1). The red straight lines are merely visual guides simulating tunnelling paths, which rigorously represent paths between the reaction-coordinate turning points for each total energy.

**Figure 4.** Arrhenius plots of the OH + CD$_4$ rate constants computed using PES-2014: CVT/LCG3, blue line; RPMD, black line. Vertical dashed line represents the crossover temperature, T$_c$ = 270 K.

**Figure 5.** CH$_4$/CD$_4$ KIEs as a function of the temperature (K) over the temperature range 200-1000 K, using the VTST/MT (blue line) and RPMD (black line) methods on the PES-2014 surface. Experimental values (green line) from Ref. 28, over the temperature range 293–800 K and from Ref. 32, over the temperature range 220–415 K.

**Figure 6.** Arrhenius plots of the OD + CH$_4$ rate constants computed using PES-2014: CVT/LCG3, blue line; RPMD, black line. Vertical dashed line represents the crossover temperature, T$_c$ = 370 K.

**Figure 7.** OH/OD KIEs as a function of the temperature (K) over the temperature range 200-1000 K, using the VTST/MT (blue line) and RPMD (black line) methods on the PES-2014 surface. Experimental values (green line) from Ref. 32, over the temperature range 220–415 K. The insert represents a zoom in the common temperature range, 250-450 K.

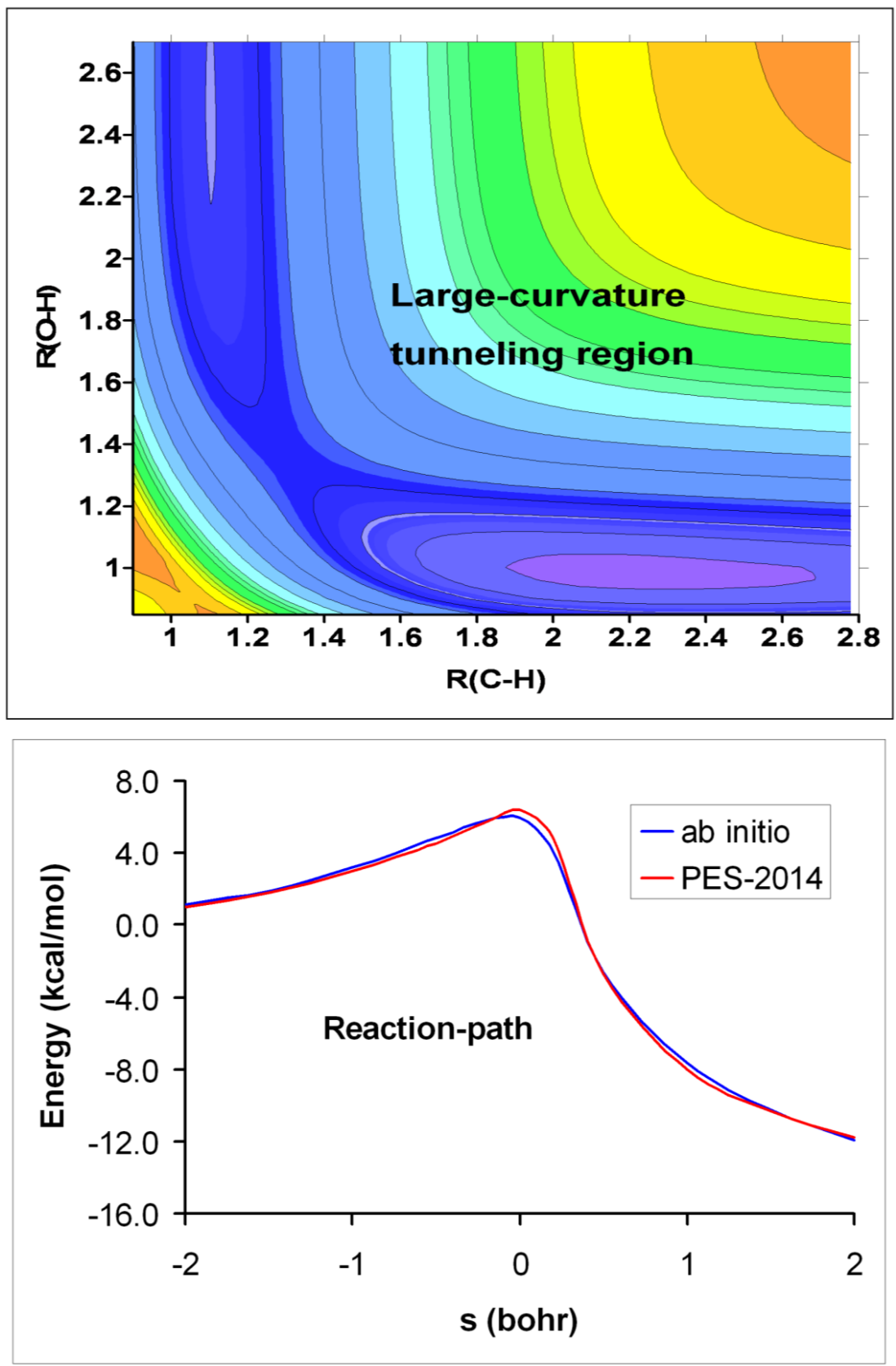

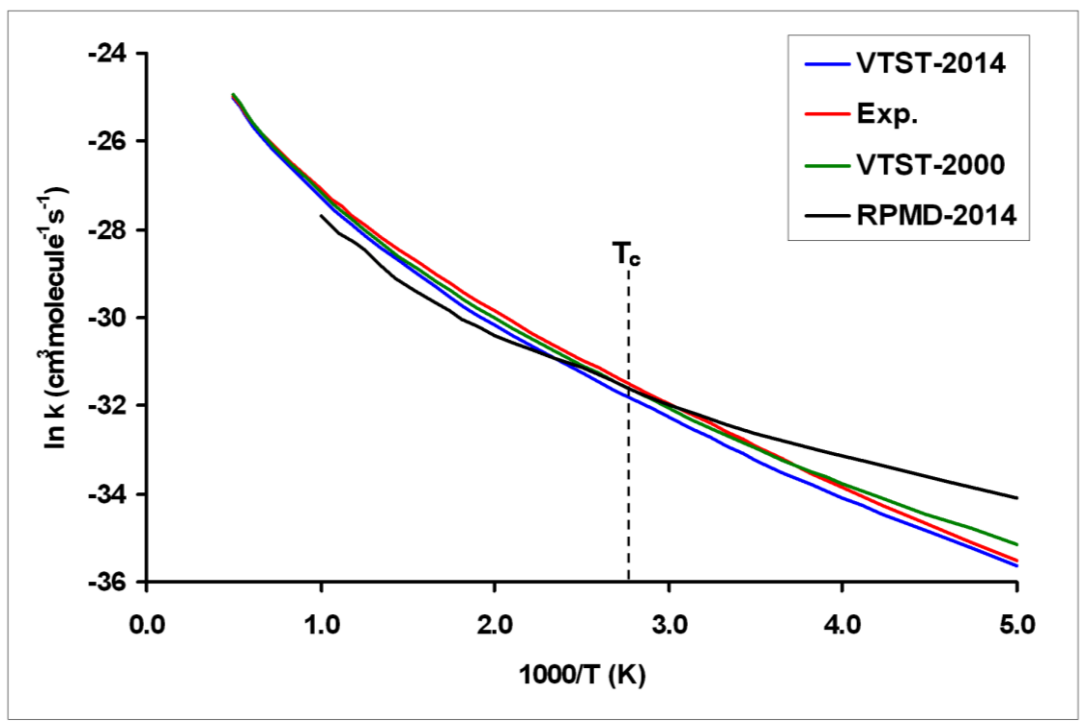

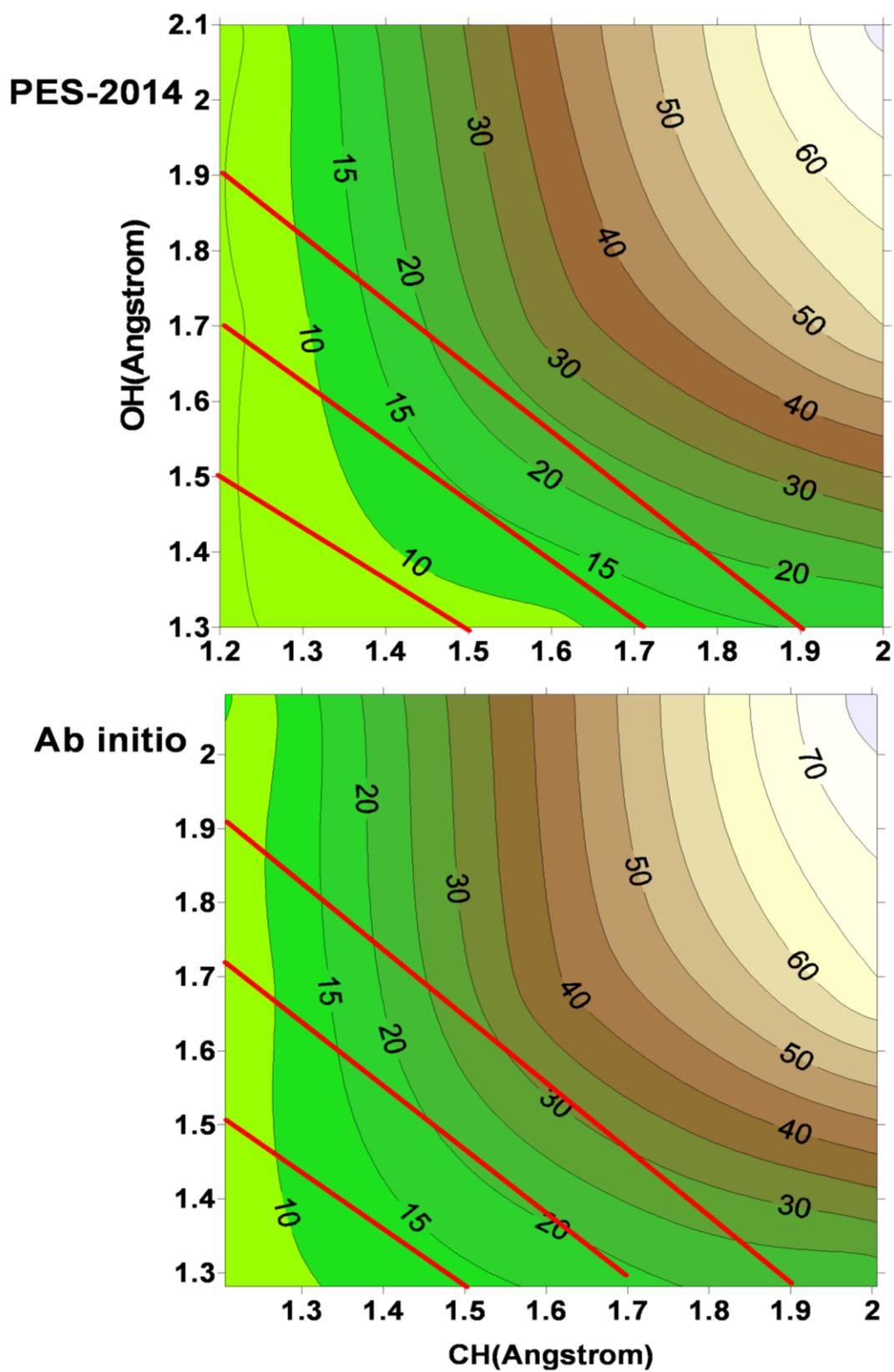

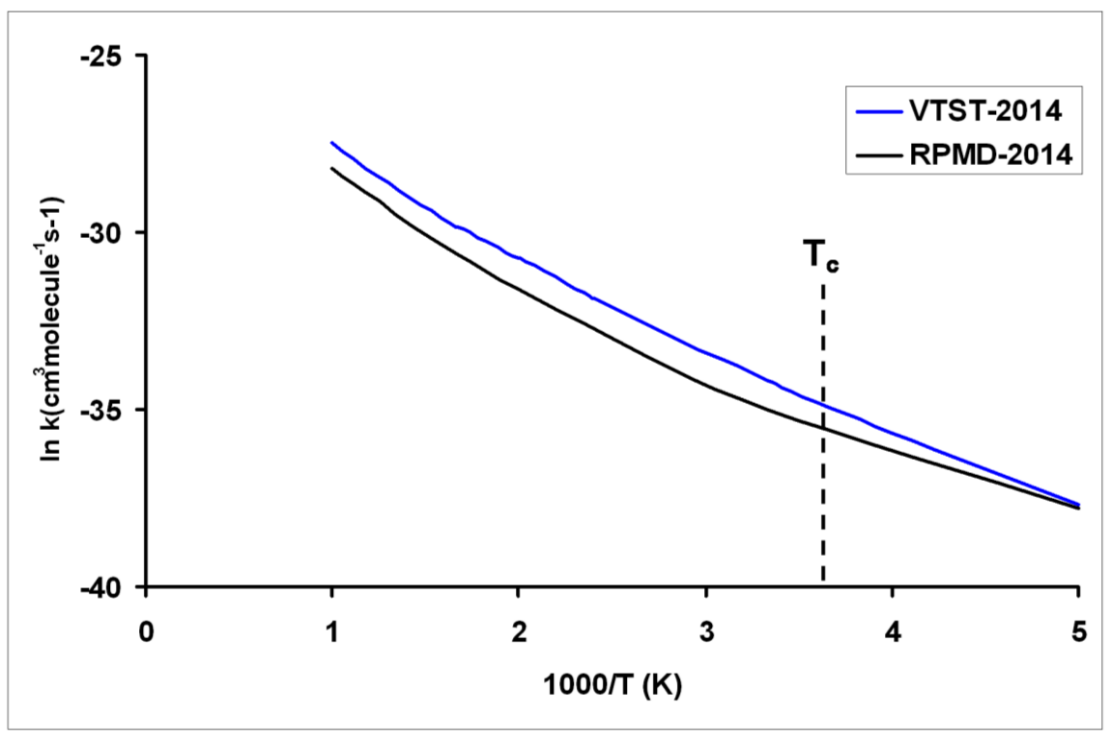

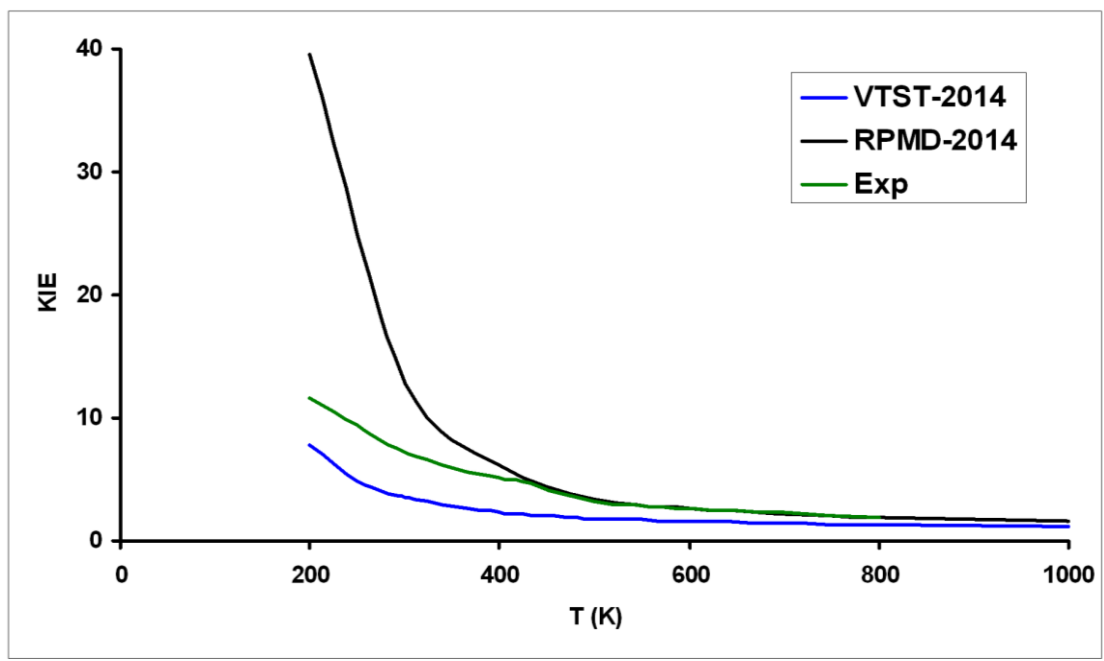

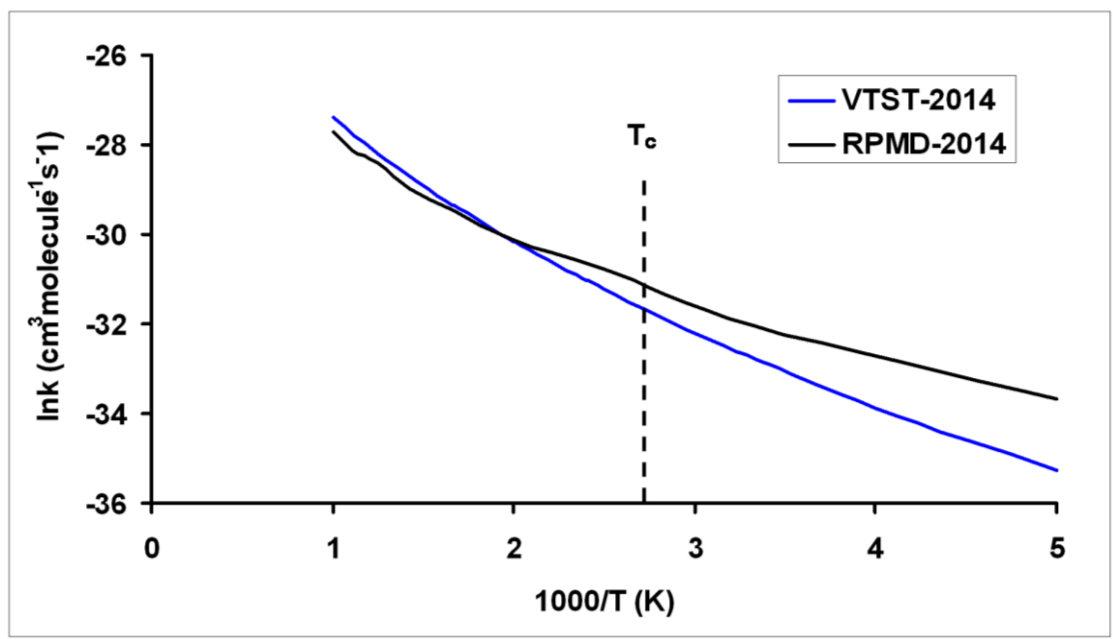

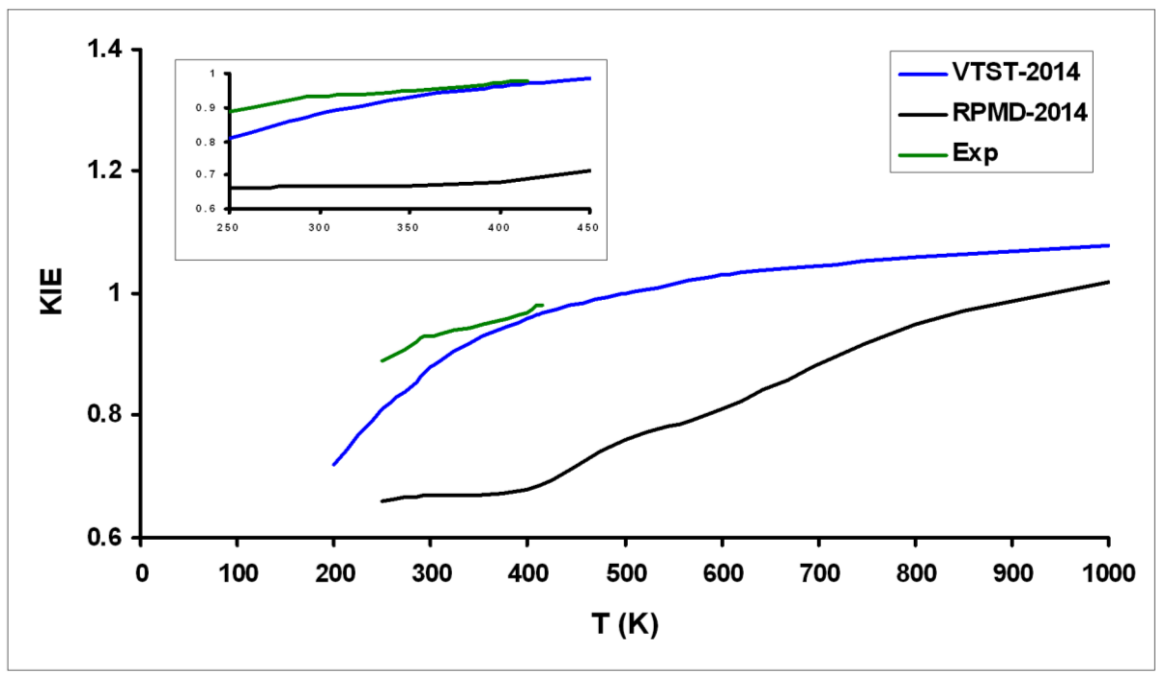